# The Impact of Artificial Intelligence on Art Research: An Analysis of Academic Productivity and Multidisciplinary Integration

Yang Ding [1,2,3*]

[1] Business School, University of Edinburgh, Edinburgh EH8 9JS, United Kingdom

[2] Edinburgh Futures Institute, University of Edinburgh, Edinburgh EH3 9EF, United Kingdom

[3] Lab for Interdisciplinary Spatial Analysis, University of Cambridge, Cambridge CB3 9EP, United Kingdom

* Correspondence concerning this article should be addressed to Yang Ding (Lab for ISA, 19 Silver Street, University of Cambridge, Cambridge CB3 9EP, email: yd386@cam.ac.uk; yang.ding@ed.ac.uk).

Abstract: This study investigates the transformative impact of artificial intelligence (AI) on art research by analysing data from 749 art research projects and 555,982 non-art research projects, as well as 23,999 journal articles. We utilized the SciBERT model for text analysis on research funding proposals and the econometric model to evaluate AI's impact on the academic productivity and impact. Our findings reveal that AI has significantly reshaped the role of art across various disciplines. The integration of AI has led to a notable expansion in keyword networks, highlighting advancements in visual art creation, data-driven methodologies, and interactive educational tools. AI has also facilitated the integration of art knowledge into nearly all research disciplines, contrasting with the traditionally confined distribution of art knowledge. Despite the substantial increase in publication impact and citation counts facilitated by AI, it has not markedly improved the likelihood of publishing in high-prestige journals. These insights illustrate the complex nature of AI's impact—enhancing research impact while presenting challenges in publication efficiency and multidisciplinary integration. The study offers a nuanced understanding of AI's role in art research and suggests directions for addressing the ongoing challenges of integrating art and AI across disciplines.

# INTRODUCTION

The rapid advancement of artificial intelligence has precipitated profound changes across numerous fields, with its effect on the domain of art research being particularly noteworthy. As AI continues to evolve, it is increasingly being integrated into creative processes, resulting in a significant transformation in how art is conceived, produced, and studied (Hemment et al., 2023; Messingschlager & Appel, 2023). This integration of AI into art research has not only introduced new tools and methodologies but has also redefined the boundaries of creativity and multidisciplinary collaboration (Huang et al., 2015; Lee et al., 2021). The convergence of AI and art, once distinct and separate domains, has fostered a new landscape where technology and creativity coalesce, challenging traditional notions of artistic expression and research paradigms (Barale, 2021; Berg, 2016; Haase et al., 2023; Hitsuwari et al., 2022).

Governments around the world have recognized the potential of AI to drive innovation across various disciplines and have increasingly supported the integration of AI into different fields of research (Zuiderwijk et al., 2021). This support is evident in the allocation of substantial funding towards multidisciplinary projects that combine AI with domains such as medicine, education, and the arts. For example, the U.S. NSF has prioritized research that leverages AI to address complex societal challenges, while the European Union's Horizon Europe program includes significant investments in AI research aimed at fostering cross-disciplinary collaborations (Huerta et al., 2022; von Krogh, 2018). These governmental efforts are designed to advance scientific and technological frontiers, while also exploring AI's potential to enhance creativity and multidisciplinary knowledge production. By fostering collaborations between AI experts and researchers in other fields, these initiatives aim to accelerate the development of innovative solutions that transcend traditional academic boundaries (Khan et al., 2023; Wagner et al., 2018).

The role of AI in art research extends beyond mere technological enhancement; it represents a paradigm shift that affects the theoretical and practical frameworks within which art is understood and practiced. Traditional art research, historically rooted in manual processes and aesthetic-focused inquiries, is increasingly intersecting with computational techniques and data-driven methodologies (Asare et al., 2023; Walczak & Moore-Pizon, 2023). The integration of AI has expanded the scope of artistic research, enabling new forms of creative expression such as generative art, where algorithms and machine learning models play a pivotal role in the creation process (O'Toole & Horvát, 2024; Stork, 2023). This shift not only alters the nature of art itself but also the academic disciplines that engage with it, leading to a more multidisciplinary approach that incorporates insights from computer science, psychology, architecture, and cognitive science (Conway, 2012; Kim et al., 2024; Rosenberg, 2016; Sauvé et al., 2022; Zaidel, 2010).

In addition to governmental support, the intersection of AI and art has garnered significant interest from both the public and private sectors, leading to increased investment and collaboration (Jankin et al., 2018; Mariani et al., 2023). Market-driven forces, including tech companies and cultural institutions, have recognized the value of integrating AI into artistic endeavors. Companies such as Google and Adobe have developed AI-powered tools that empower artists to explore new creative possibilities, while cultural organizations have initiated programs to support artists in experimenting with AI, often providing grants, residencies, and exhibition opportunities (World Economic Forum, 2024). This confluence of government policy and market investment is creating a fertile environment for the growth of AI-enhanced art, driving both innovation and public engagement. By combining resources from both sectors, these initiatives not only expand the reach of AI within the arts but also contribute to the broader dialogue about the future of creativity in the digital age (Messer, 2024; Walczak & Moore-Pizon, 2023).

The intersection of AI and art has also begun to exert significant effect on other academic disciplines, fostering a broader multidisciplinary exchange and contributing to knowledge diversity. AI-enhanced artistic techniques have been employed in biology to create visualizations that aid in understanding complex biological processes, improving both scientific communication and education (Fan & Zhong, 2022). In the social sciences, AI-generated art has been used to explore cultural trends and social dynamics, offering new perspectives on societal issues (von Krogh, 2018). The integration of AI and art in these contexts not only enriches the methodological toolkit of these disciplines but also encourages a more holistic approach to research, where creativity and technology are leveraged together to tackle complex problems. This multidisciplinary impact underscores the importance of considering AI and art not just as isolated phenomena but as catalysts for broader academic and societal transformation (Adam, 2023; Kalpokiene & Kalpokas, 2023).

To understand the impact of artificial intelligence (AI) on art research, this study utilizes a dataset comprising 749 NSF-funded art-related research projects and 555,982 non-art-related research projects across 12 NSF directorates and 39 disciplinary areas. We analyze keyword networks from proposal abstracts of traditional art projects compared to those of AI-enhanced art projects to quantify the impact of AI on the evolution of art knowledge. Additionally, we evaluate the similarity between research proposals from traditional art projects, AI art projects, and other disciplines. By treating AI as a variable, we assess its impact on the scientific productivity and research impact of scholars involved in art projects. Our findings reveal that AI has significantly impacted the scope and dissemination of art research, leading to advancements in visual technologies and interactive tools. However, despite these advancements, AI has not substantially increased the likelihood of publishing in high-prestige journals, highlighting a complex interplay between multidisciplinary innovation and academic prestige.

The remainder of this paper is organized as follows: Section 2 provides a comprehensive review of the theoretical background, including the role and impact of AI in artistic practices and the multidisciplinary nexus of AI and art. Section 3 shows the data used in this study. Section 4 details the methodology. Section 5 presents the results of our analysis. Section 6 concludes with a summary of the key insights.

# THE CONVERGENCE OF ART AND ARTIFICIAL INTELLIGENCE: THEORETICAL BACKGROUND

The role and impact of AI in artistic practices

Artificial intelligence has fundamentally transformed artistic practices by introducing new tools and techniques that redefine the boundaries of creativity. Recent advancements, such as those demonstrated by deep learning models like Dall-E 2 and Midjourney, have enabled artists to generate high-quality artworks based on textual descriptions or style emulation (Oksanen et al., 2023; Zhou & Lee, 2024). These AI tools allow for the exploration of novel aesthetic forms, combining diverse artistic genres in ways previously unattainable (Ornes, 2019). The ability of AI to generate new art by analyzing vast datasets of existing works has led to innovative artistic outputs that reflect a fusion of multiple styles, pushing the limits of traditional art creation (Chatterjee, 2022).

AI's impact extends beyond the creative process to enhance audience engagement and interaction. Interactive installations powered by AI, which use real-time data from sensors and cameras, create immersive experiences in museums and galleries (Chatterjee, 2022; Zylinska, 2023). These installations adapt to the behavior of viewers, offering dynamic and personalized interactions that transform static art displays into engaging, responsive environments. This shift represents a significant evolution from conventional art forms, highlighting AI's role in creating more fluid and interactive artistic encounters that adapt to audience inputs (Latikka et al., 2023).

The integration of AI into artistic practices also raises important questions about authorship, creativity, and intellectual property. The ability of AI to generate artworks that closely mimic or merge different styles has sparked debates about the originality and value of AI-created art versus human-created art (Jr et al., 2023; Then et al., 2023). Concerns about AI eroding the role of the artist and the uniqueness of human creativity are prominent, with critics arguing that the widespread use of AI in art could diminish the perceived value of human artistic effort (Jiang et al., 2023). These debates reflect broader societal concerns about the implications of automation and AI in creative fields, emphasizing the need for thoughtful consideration of how these technologies affect traditional notions of creativity and artistic worth (Then et al., 2023).

The multidisciplinary nature of AI and art research underscores the complexity of these issues, blending insights from technology, philosophy, and art theory. Historical perspectives on non-human-made art and ongoing philosophical debates about creativity and authorship are crucial for understanding AI's role in contemporary art (Oksanen et al., 2023; Zylinska, 2023). As AI continues to evolve, it is essential to address the ethical and practical challenges it presents, including intellectual property concerns and the changing nature of artistic practice (Then et al., 2023). This intersection of disciplines highlights the need for comprehensive approaches to fully grasp the impact of AI on the future of art and creativity.

The multidisciplinary nexus of AI and art

The integration of Artificial Intelligence with art represents a burgeoning multidisciplinary domain, combining computational advancements with creative expression. This synthesis offers numerous advantages, primarily through the deployment of sophisticated algorithms such as Generative Adversarial Networks (GANs) and deep learning frameworks. These technologies have revolutionized artistic practices by enabling the generation of novel and diverse art forms that transcend traditional boundaries (Goodfellow et al., 2014; Karras et al., 2017). AI facilitates the exploration of new creative avenues, enhancing both the quantity and variety of artistic outputs. For instance, GANs have been instrumental in producing high-quality, unique visual content, which was previously unattainable through conventional methods (Karras et al., 2017).

The confluence of AI and art is not confined to the realm of creative practice but extends its impact across various scientific disciplines. In the field of biology, the concept of BioArt has emerged, wherein AI is utilized to visualize and interpret biological phenomena in innovative ways (Simou et al., 2013; Yetisen et al., 2015). This multidisciplinary approach not only enhances our understanding of biological systems but also stimulates advancements in synthetic biology (Aithani et al., 2023; Baranzini et al., 2022; Kac, 2020). Similarly, in medical imaging, AI-driven art tools are employed to enhance the visualization and interpretation of complex medical data, thereby facilitating more accurate diagnostics and treatment planning (Huston & Kaminski, 2023; Potier, 2011; Shen et al., 2017; Yale Medicine Magazine, 2005). Furthermore, in computer science, the integration of AI in artistic contexts has propelled advancements in computer vision and image processing, fostering novel methodologies for image generation and style transfer (Isola et al., 2016; Ma & Huo, 2024).

The rapid pace of innovation facilitated by AI in art has sparked debate regarding the speed and depth of knowledge discovery. Proponents argue that AI's capability to process and analyze vast datasets expedites the creation and experimentation processes, leading to accelerated advancements in both artistic and scientific domains (Epstein et al., 2023; Messer, 2024). This swift generation of novel content allows for extensive exploration of creative possibilities. However, critics contend that such rapidity may lead to superficial or transient innovations, lacking the depth and substantive impact of more traditionally developed artistic and scientific insights (Epstein et al., 2023). This discourse highlights the tension between the efficiency of AI-driven processes and the quality of the resultant knowledge, underscoring the need for a balanced evaluation of both speed and substance in multidisciplinary research (Adam, 2023).

## DATA

This study utilizes a comprehensive dataset comprising 560,149 research projects funded by the National Science Foundation (NSF) from 1955 to 2024. Established by the U.S. federal government, the NSF is a key agency dedicated to supporting basic scientific research, fostering technological innovation, and advancing science education. This dataset provides a thorough basis for analyzing the role of art within the broader context of NSF-funded research, highlighting its multidisciplinary significance and contributions.

Among the dataset, 749 projects are identified as art-related, covering 39 disciplinary areas across 12 NSF directorates. These directorates include the Directorate for Biological Sciences (BIO), Directorate for Computer and Information Science and Engineering (CISE), Directorate for Education and Human Resources (EHR), Directorate for Engineering (ENG), Directorate for Geosciences (GEO), Directorate for Mathematical and Physical Sciences (MPS), Office of Integrative Activities (OIA), Office of International Science and Engineering (OISE), Directorate for Research, Innovation, Synergy, and Education (RISE), Directorate for Social, Behavioral and Economic Sciences (SBE), Directorate for Technology Innovation (TI), and Directorate for Environmental Engineering and Science Excellence and Equity (EES). Within these art-related projects, 32 specifically investigate the integration of art with artificial intelligence (AI), reflecting the NSF's support for innovative, multidisciplinary research at the intersection of art and emerging technologies.

In addition to the art-related projects, the dataset includes 555,982 non-art-related research projects. These projects cover a broad spectrum of academic fields, providing a comprehensive context for comparing the semantic similarities and multidisciplinary connections between art-related and non-art-related research.

The dataset encompasses detailed information for each project, including the names and affiliations of principal investigators (PIs), contract numbers, funding start and end dates, awarded amounts, and research proposal abstracts. Gender information for PIs is also included, offering additional insights into the demographic aspects of the funded research. The distribution of art-related research projects, AI-art integration projects, and other research funding projects across the 12 NSF directorates and 39 disciplinary areas is detailed in **Supplementary Note 1**.

We obtained 23,999 journal articles published by principal investigators of art-related projects from the MAG database. The metadata includes titles, abstracts, affiliations, publication year, citation counts, journal CiteScore, and the research fields of the articles. We compiled statistics on the number of articles and citation counts for the scholars' respective research fields and institutions. The MAG database, developed by Microsoft Research, is a comprehensive academic resource offering unique advantages for scholarly research. MAG covers diverse academic fields, including science, technology, engineering, medicine, humanities, and arts, providing a multidisciplinary platform rich in cross-disciplinary resources. Utilizing advanced data mining and machine learning technologies, MAG efficiently collects, processes, and intelligently links authors, institutions, journals, and citation information, creating a high-quality academic network. The database is dynamically updated, ensuring the timeliness and accuracy of data. MAG provides detailed metadata, such as article titles, authors, publication years, journal names, citation counts, and CiteScore, supporting in-depth literature analysis and academic evaluation. Its robust citation aids in analyzing academic impact and research trends and tracking knowledge dissemination.

## METHODS

Text embedding and dimensionality reduction

In this study, we utilize the SciBERT model to embed abstract texts and capture high-dimensional semantic features. Developed by the Allen Institute for AI, SciBERT is a pre-trained language model specifically designed for the scientific literature domain (Beltagy et al., 2019). SciBERT, having been pre-trained on a large corpus of scientific literature, demonstrates superior comprehension of scientific terminology and complex syntactic structures, thereby generating high-quality semantic embeddings for scientific document abstracts.

Prior to embedding the texts, we performed preprocessing steps including converting texts to lowercase, removing punctuation, tokenization, filtering out stop words and performing lemmatization. These preprocessing steps transformed the texts into a format suitable for input into the SciBERT model, resulting in high-dimensional embedding vectors. To further reduce the dimensionality of these embedding vectors and facilitate visualization, we employed Uniform Manifold Approximation and Projection (UMAP) technology (McInnes et al., 2018). UMAP is a popular nonlinear dimensionality reduction technique known for its efficiency in handling high-dimensional data while preserving both global and local structures. By reducing dimensions to a two-dimensional space with UMAP, we are able to more intuitively observe and analyze the distribution of text embeddings, providing support for subsequent clustering analysis and visualization.

Keyword extraction and clustering

Following text embedding and dimensionality reduction, we applied the Term Frequency-Inverse Document Frequency (TF-IDF) method for keyword extraction from the preprocessed texts (Grootendorst, 2022). TF-IDF is a commonly used weighting method in information retrieval and text mining, measuring the importance of words within a document collection. By calculating the TF-IDF values for each term, we can extract significant keywords from each abstract and analyze these keywords based on their weights.

For clustering text abstracts, we used the K-means++ clustering algorithm (Arthur & Vassilvitskii, 2007) . K-means++ is a classic clustering method known for its simplicity, efficiency, and rapid convergence, making it particularly suitable for large-scale data clustering. In this study, we performed K-means++ clustering on the reduced-dimensional embedding vectors, specifying various numbers of clusters to reveal the latent thematic structures within the text abstracts. After clustering, we visualized the results and provided semantic interpretations for each cluster based on the TF-IDF extracted keywords.

Semantic similarity calculation of research proposals

In this study, we calculate two primary similarity metrics to quantify the semantic similarity between art-related research grant proposals and research proposals from other disciplines funded by the NSF: weighted similarity and max similarity. These metrics are designed to enhance the robustness and comprehensiveness of our analysis.

The weighted similarity metric measures the overall similarity between an art project's proposal abstract and the proposal abstracts from other disciplines. To compute weighted similarity, we first calculate the cosine similarity between the art project and each discipline project. Then, these similarities are weighted, with the weights being inversely proportional to the number of projects in the corresponding the discipline. The use of the inverse of the project count as a weight ensures that disciplines with fewer projects are not underrepresented in the weighted similarity calculation. The equation for calculating weighted similarity is as follows:

$$\text{Weighted similarity}_i = \sum_{j=1}^{m} \frac{1}{N_j} \times C(A_i, B_j) \tag{1}$$

where $A_i$ represents the three-dimensional tensor of the $i$th art project's proposal abstract. $B_i$ represents the three-dimensional tensor of the $i$th project's proposal abstract in a discipline other than that of $A_i$. $N_j$ is the total number of projects in the other discipline containing project $B_j$. The $m$ is the total number of projects in disciplines other than the one to which $A_i$ belongs. $C(A_i, B_j)$ is the cosine similarity of $A_i$ and $B_i$.

The max similarity metric identifies the discipline, other than the one to which the art project $A_i$ belongs, that has the highest weighted similarity with the art project. The max similarity for the $i$th art project is:

$$\text{Max similarity}_i = \max_{k \neq Art} \left( \frac{1}{N_k} \sum_{j=1}^{m_k} \times C(A_i, B_j) \right) \tag{2}$$

where $N_k$ is the total number of projects in the $k$th discipline. The $m_k$ is the number of projects in the $k$th discipline.

To visually represent the relationships between clusters and keywords, we constructed keyword networks. Separate networks were established for abstracts of traditional art projects and combined art-AI projects to observe distinctions. Additionally, to examine the impact of AI on the aggregation of art knowledge with different NSF disciplines, we performed within-discipline clustering of all NSF-funded research abstracts. Furthermore, we constructed keyword networks for clusters of traditional art projects and combined art-AI research funding projects, comparing them with clusters from other disciplines. In the network graph, nodes from different clusters are color-coded, and node sizes are determined by their TF-IDF values, providing a clearer visualization of each keyword's importance. We calculated the cosine similarity of knowledge between research proposals to assess the impact of AI on the dissemination of art knowledge.

Econometric model

To account for individual fixed effects of scholars and time-varying factors affecting productivity, it is beneficial to incorporate dummy variables for *AI*, *Art*, and their interaction terms within the model. This approach typically involves using a two-way fixed effects (TWFE) model (De Chaisemartin & D'Haultfœuille, 2020; Egami & Yamauchi, 2021). By doing so, we can effectively isolate the impact of specific interventions or changes over time while controlling for unobserved heterogeneity among individuals and temporal trends. The foundational model for this analysis is as follows:

$$Y_{it} = \beta_0 + \beta_1 AI_{it} \cdot Art_t + \beta_2 AI_{it} + \beta_3 ART_t + \delta X_{it} + \omega_i + v_t + \varepsilon_{it} \qquad (3)$$

where $Y_{it}$ is the academic performance of scholar $i$ at time $t$; $AI_{it} \cdot Art_t$ is the interaction term between AI and funded art projects, capturing the differential impact of AI when scholars are involved in funded art projects. $Art_t$ is a dummy variable for the time period $t$, representing the presence of a funded art project. $X_{it}$ is the vector of covariates. $\omega_i$ and $v_t$ indicate scholars' fixed effects and time fixed effects, respectively. $\beta_1$ is the coefficient of interest.

To control for various external factors that may affect academic productivity, the model includes several key covariates. These include the amount of research funding received, the academic age of scholars, and their gender. The model also incorporates metrics related to the research fields of the scholars, such as the total number of publications and citations within each field. Institutional factors are accounted for as well, including the total number of publications and citations for the scholar's affiliated university or college, the amount of ART-related funding and training at the scholar's institution, and the historical number of art-related grants awarded to the institution (Baccini et al., 2014; Yegros-Yegros et al., 2015). These covariates help control for external impacts and provide a clearer understanding of the impact of AI on funded art projects.

To address potential reverse causality, the model incorporates scholars' initial academic performance (Lawson et al., 2021). This includes their publication count, citation count, and CiteScore for the three years preceding the study period, with dummy variables for these years representing their baseline academic output. This approach ensures that the effects of AI and funded art projects on productivity are not confounded by scholars' prior performance, allowing for a more accurate assessment of how AI affects productivity in the context of funded art projects.

This study utilizes annual scientific publication counts, citation counts, and CiteScore of publications as primary metrics for evaluating academic performance among scholars in the arts. Each of these metrics offers distinct and complementary insights into the scholarly impact and quality of research outputs. Annual publication counts measure the frequency and volume of research contributions, providing a straightforward indication of productivity (Azoulay et al., 2021). Citation counts reflect the academic impact of these publications, revealing how often they are referenced by other researchers and their significance within the field (Jacob & Lefgren, 2011). CiteScore provides a measure of the quality and reputation of publication venues. Compared to the Journal Impact Factor (JIF), it relies on a four-year citation window, yielding a more comprehensive assessment of citation impact. It also encompasses a broader range of journal types, mitigating the bias towards journals with higher citation frequencies and delivering a more inclusive perspective on journal performance (Fernandez-Llimos, 2018).

We also assess whether AI contributes to broadening the scope of art research by enhancing the integration of art knowledge with other scientific and technological fields. To evaluate this, we use Equation 3 once more to estimate the impact of AI on the multidisciplinary semantic similarity metrics of research proposals. Here, weighted similarity and max similarity are used as the dependent variables, with AI serving as the treatment variable.

# RESULTS

AI-driven expansion in the diffusion of art research across academic disciplines

The introduction of AI into traditional art research appears to be influencing the way knowledge is disseminated across various academic disciplines. This impact can be observed through the analysis of the knowledge networks shown in the provided figure, which highlights changes in the integration and dissemination patterns before and after AI's incorporation into art-related research.

The dissemination of traditional art knowledge across disciplines shows a foundational level of multidisciplinary exchange, yet it remains somewhat localized and constrained. As illustrated in Fig. 1(a), the knowledge network of NSF-funded traditional art research projects includes red nodes that represent key topics such as "creativity," "design," and "undergraduate," all of which are integral to education and the creative process in traditional art. These nodes are connected to other research domains (blue nodes), which shows that traditional art knowledge does extend beyond its core discipline. For example, "creativity" is linked with education, highlighting the role of art in enhancing cognitive development and innovative thinking in educational contexts. However, the connections between these nodes are relatively sparse, and the clusters are somewhat isolated, suggesting that while art knowledge does disseminate to other fields, it tends to remain within specific, and sometimes narrowly defined, boundaries.

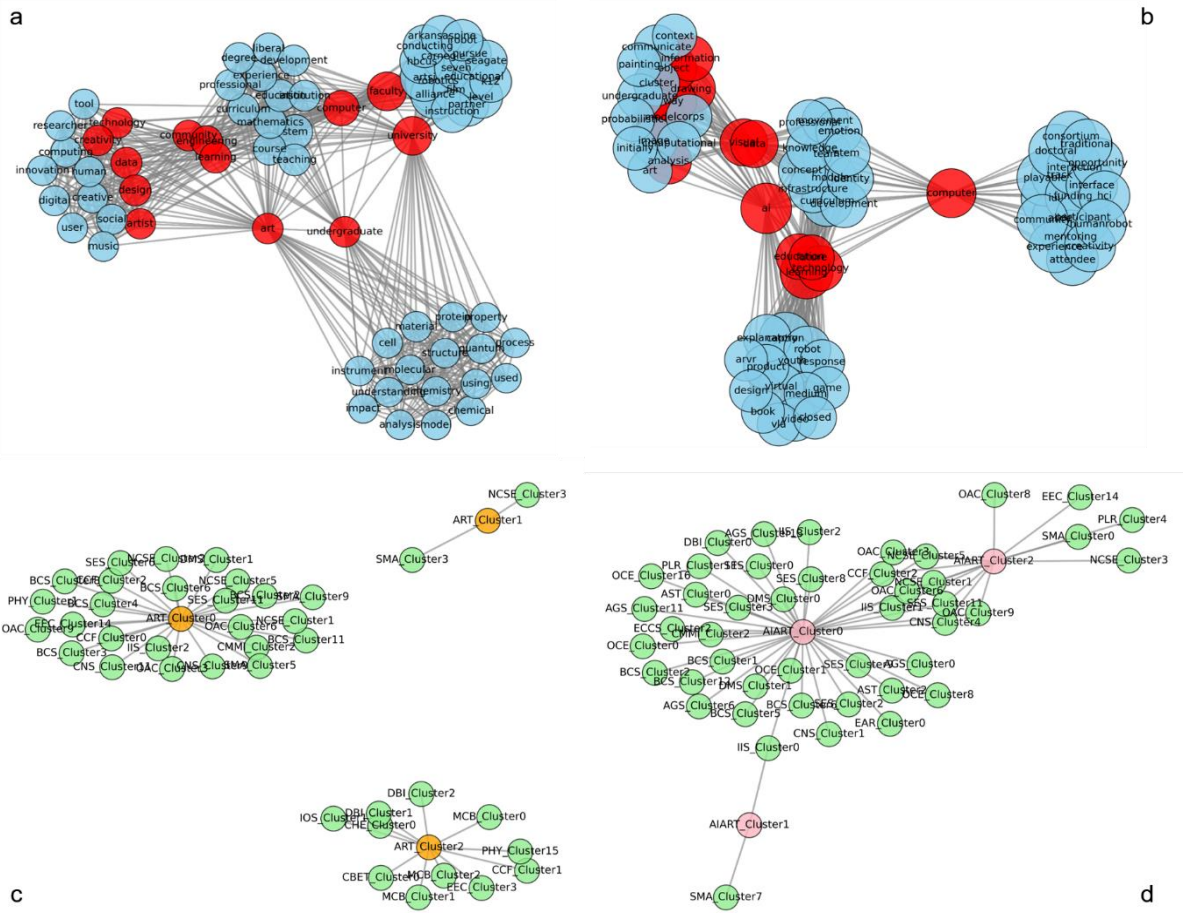

**Fig. 1 Evolution of Knowledge Networks in Traditional Art and AI-Enhanced Art Research.** Fig.1(a) shows the knowledge networks in traditional art research. Fig.1(b) illustrates the knowledge networks in AI-enhanced art research. Fig.1(c) depicts the knowledge networks of clusters from traditional art projects compared with clusters from other NSF disciplines. Fig.1(d) presents the knowledge networks of clusters from AI-enhanced art projects compared with clusters from other NSF disciplines.

The integration of AI with traditional art knowledge has the potential to significantly increase the interconnectedness and density of knowledge networks. In Fig. 1(b), the introduction of AI-related concepts (such as "machine learning," "data," and "intelligence") into the art domain transforms the knowledge network into a more interconnected and potentially more influential structure. The red nodes in this AI-enhanced network, representing AI-art intersections, now hold central positions and are connected to a broader range of scientific fields. This shift implies that AI could be amplifying the reach of art knowledge, facilitating the cross-pollination of ideas between traditionally separate domains. The AI-art nodes reflect a shift towards more data-driven and technologically infused artistic processes, such as using machine learning to analyze artistic styles or employing AI to create generative art. This increased interconnectedness suggests that AI might be enabling new forms of creativity that not only affect art itself but also contribute to other areas like computational sciences and cognitive studies. Nonetheless, it is important to consider that this enhanced connectivity represents an ongoing evolution, with the full impact of AI's integration still unfolding.

The aggregation of knowledge and its cross-disciplinary integration is enhanced by AI, which fosters more extensive interactions between art-related research and other scientific fields. Fig. 1(c) and (d) demonstrate how the introduction of AI into art research can alter the degree to which art knowledge aggregates and integrates across different disciplines. Fig. 1(c) shows that traditional art clusters (green nodes) have a certain degree of interaction with NSF-funded clusters in other fields, yet these interactions appear limited. The traditional art clusters are somewhat isolated, which may indicate that while art knowledge interacts with other disciplines, it primarily remains within its own or closely related areas. In contrast, Fig. 1(d), representing the AI-art network, shows a noticeable increase in connectivity. The AI-art clusters (orange and red nodes) are much more integrated with clusters from a variety of other NSF disciplines. This increased connectivity could suggest that AI is not only helping to aggregate art knowledge more effectively but is also enabling it to permeate into a broader spectrum of scientific discussions, including those traditionally distant from the arts. The growth from 239 connections in the traditional art network to 261 in the AI-art network highlights this potential expansion, though the extent and depth of these new connections require further analysis. The 263 clusters, which represent all research funding projects related to arts, the integration of arts with AI, and other NSF disciplines, are detailed in **Supplementary Note 2**. This note also includes the full data on the nodes, edges, and weights derived from the abstracts of NSF-funded projects combining AI with the arts, as well as those focused on traditional arts.

The implications of AI on the cross-disciplinary dissemination and aggregation of art knowledge are multifaceted, offering both opportunities and challenges for research and policy. The evolving knowledge networks suggest that AI could be facilitating the emergence of new multidisciplinary fields that combine the creative aspects of art with the technical advancements of AI. These new intersections may lead to innovative research methodologies and applications that could transcend traditional disciplinary boundaries. Funding agencies and policymakers may need to recognize the importance of these AI-art intersections in driving cross-disciplinary knowledge exchange and consider supporting initiatives that foster such collaborations. Furthermore, academic institutions may find it necessary to gradually adapt their educational and research frameworks to accommodate the increasing interdisciplinarity fostered by AI-art synergies. However, these implications should be considered as part of a broader, ongoing process, with the understanding that the full impact of these shifts will become clearer as the fields continue to evolve and integrate.

Following the network graph analysis, which illustrated the patterns of multidisciplinary in art projects, our quantitative and model-based analysis further validates these findings. Our regression results indicate that AI integration significantly boosts the multidisciplinary of art projects, as show in Fig. 2. The coefficient for AI integration is 0.507 ($p<0.001$) for weighted similarity, which measures the average similarity between art projects and all other disciplines, and 0.259 ($p<0.001$) for maximum similarity, which measures the highest similarity with a single other discipline. The more substantial positive effect observed for weighted similarity suggests that AI broadly enhances the overall multidisciplinary of projects by facilitating the infusion of diverse disciplinary knowledge. This effect is nearly twice as large as the impact on maximum similarity, highlighting AI's broader impact on cross-disciplinary engagement.

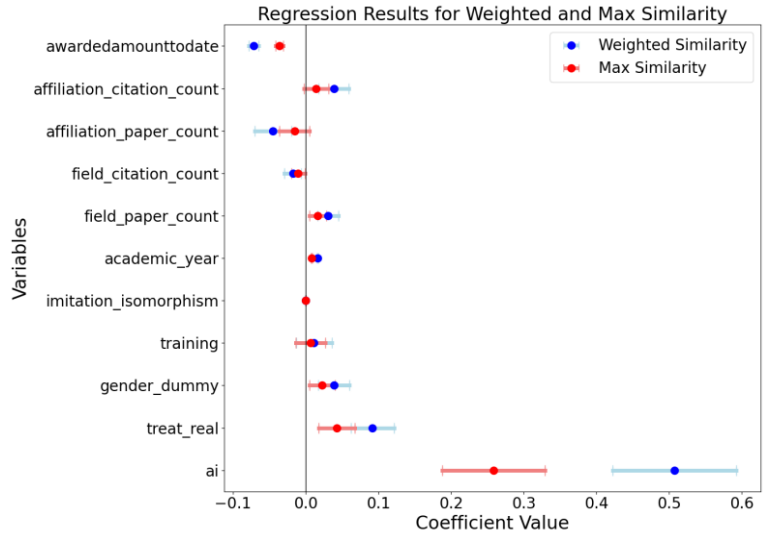

**Fig. 2 Results of the impact of the integration of artificial intelligence and art on the multidisciplinary of knowledge.**

Research funding also plays a critical role in increasing multidisciplinary. The variable treat_real, which indicates whether a project received financial support, has coefficients of 0.092 (p < 0.001) for weighted similarity and 0.043 (p<0.001) for maximum similarity. Funded projects tend to engage with a wider range of disciplines, likely due to the enhanced resources that support extensive collaborations. As with AI integration, the effect of funding on weighted similarity is more than double its effect on maximum similarity, underscoring that financial support generally enhances the average level of multidisciplinary engagement rather than just the peak.

In contrast, institutional affiliation and academic output show marginal contributions to multidisciplinary. For instance, affiliation citation count and affiliation paper count have coefficients of 0.039 (p<0.001) and -0.045 (p<0.001) for weighted similarity, and 0.014 (p = 0.099) and -0.015 (p=0.147) for maximum similarity. Similarly, field citation count and field paper count display coefficients of -0.018 (p=0.003) and 0.030 (p<0.001) for weighted similarity, and -0.010 (p=0.038) and 0.017 (p=0.006) for maximum similarity. These results suggest that while higher academic output and citation impact may slightly favor multidisciplinary integration, their effects are less pronounced compared to AI and funding.

Temporal factors and individual characteristics exhibit limited effect on multidisciplinary. The variable academic year shows minor but statistically significant coefficients of 0.016 (p< 0.001) for weighted similarity and 0.008 (p<0.001) for maximum similarity. Conversely, variables such as imitation isomorphism, training, and gender dummy show negligible effects, with coefficients and p-values indicating limited impact on multidisciplinary. The complete regression results can be found in **Supplementary Note 3**.

The impact of AI integration on publication efficiency and impact in art projects

Table 1, Column 1, shows that the introduction of AI did not significantly affect research productivity among scholars in the arts compared to those not using AI. The multidisciplinary nature of combining AI with the arts may lead to additional knowledge acquisition and learning burdens. Cross-disciplinary research often requires scholars to integrate complex knowledge from two fields, resulting in extra learning and adaptation costs. Despite AI's potential to enhance research innovation, scholars may face challenges such as understanding and applying AI technology and implementing these technologies within the arts. These additional burdens could undermine AI's advantages in improving research efficiency and quality, resulting in less than expected overall improvements in academic outcomes.

**Table 1. TWFE regression estimates of AI integration effects on academic performance.**

| Variable | (1) Publication | (2) Citation | (3) CiteScore |
|---|---|---|---|
| AI × Art | 0.568 | 211.304** | 0.489 |
|  | (0.611) | (80.675) | (0.967) |
| Amount | 0.023 | -1.908 | 0.005 |
|  | (0.012) | (1.619) | (0.019) |
| Initial performance | 2.534*** | 20.010*** | 1.780*** |
|  | (0.130) | (4.489) | (0.105) |
| Career | 0.008*** | 0.842*** | 0.023*** |
|  | (0.001) | (0.138) | (0.002) |
| Gender | 0.112** | 19.286*** | 0.235*** |
|  | (0.041) | (5.463) | (0.066) |
| Field citation | 0.070** | 14.870*** | 0.177*** |
|  | (0.022) | (2.935) | (0.035) |
| Field paper | -0.028 | -16.626*** | -0.182*** |
|  | (0.027) | (3.586) | (0.043) |
| Affiliation paper | -0.227*** | -28.989*** | -0.533*** |
|  | (0.053) | (6.938) | (0.083) |
| Affiliation citation | 0.207*** | 24.414*** | 0.460*** |
|  | (0.042) | (5.578) | (0.067) |
| Training | -0.327*** | 6.466 | -0.116 |
|  | (0.051) | (6.724) | (0.081) |
| Imitation | -0.070*** | -5.383** | -0.096*** |
|  | (0.013) | (1.713) | (0.021) |
| Constant | -1.713*** | -6.135 | 2.198*** |
|  | (0.165) | (21.807) | (0.261) |
| R-squared | 0.191 | 0.148 | 0.156 |

**Note:** *** $p<0.001$, ** $p<0.01$, * $p<0.05$. Initial productivity represents the initial (Publication counts | Citation counts | CiteScore) productivity.

AI significantly increased the impact of publications for scholars in funded art projects. Column 2 of Table 1 shows that the interaction term between AI and art projects has a coefficient of 211.304. This significant effect indicates that the introduction of AI substantially enhances the impact of art scholars' publications compared to traditional art research. This increase can be attributed to several factors. AI introduces knowledge diversity by integrating advanced technologies and data analysis methods from various fields, fostering innovation and diversity in art research. Secondly, the popularity of AI has made its application a prominent research trend, attracting considerable academic attention and discussion, which naturally boosts the impact of related work. AI's advantages, such as efficient data processing and deep learning capabilities, provide artists with unprecedented insights and inspiration, leading to highly impact research outcomes. These factors collectively contribute to AI's notable role in enhancing the academic impact of art research.

AI's integration did not result in publications by art scholars appearing in higher-prestige journals. The interaction term between AI and art projects had a statistically insignificant effect on CiteScore. Although AI significantly boosts the impact of art scholars' work, it does not significantly enhance the likelihood of publishing in high-CiteScore journals. High-prestige journals often take a cautious stance towards multidisciplinary research, which may restrict the publication opportunities for AI-enhanced art research (Park et al., 2023).

## CONCLUSIONS

This study explores the impact of artificial intelligence on art research, drawing on data from the NSF and the MAG. We examined how AI has impacted the dissemination of art knowledge across various academic disciplines and its effects on the productivity and impact of art researchers. Our analysis involved 749 NSF-funded art projects and 555,982 non-art projects. We began by comparing keyword networks from traditional art projects with those from AI-enhanced art projects. We then assessed the semantic similarity of research proposals between traditional art, AI-enhanced art, and 39 non-art disciplines. To evaluate the role of AI in integrating art knowledge across disciplines, we employed a Two-Way Fixed Effects model. Additionally, we analyzed 23,999 journal articles published by researchers involved in both traditional and AI-enhanced art projects. This analysis, conducted through the TWFE model, helped us compare publication quantity, citation counts, and CiteScore between scholars in AI-enhanced art research and those in traditional art fields.

Our findings reveal a significant shift in the landscape of art research with the advent of AI. Traditional art research, previously characterized by a focus on aesthetic and manual processes with limited multidisciplinary integration, has expanded dramatically through AI. The introduction of AI-related keywords such as "data," "computational," and "digital" highlights how AI is enhancing visual art creation and facilitating innovative methodologies like generative art. This shift underscores AI's role in broadening the scope of artistic research and fostering dynamic interactions between art and technology.

The integration of AI has also led to a notable expansion of art's effect across various academic disciplines. Whereas traditional art research was often confined to aesthetic and educational contributions, AI has enabled art to permeate new areas such as scientific visualization and technological innovation. Our keyword network analysis shows an increased presence of terms like "technology" and "visual," reflecting AI's role in enhancing scientific communication and creating interactive educational tools. An important finding of this study is that AI has enabled art knowledge to permeate nearly all disciplines and clusters within the NSF's research portfolio. This contrasts with the traditional distribution of art knowledge across various disciplinary research proposals, which was relatively confined. The integration of art across diverse fields, facilitated by AI, demonstrates a more extensive and interconnected role for art in multidisciplinary research. This expanded role is further supported by our economic modeling results, which indicate that AI encourages scholars to incorporate more cross-disciplinary knowledge into their research proposals.

Despite these advancements, AI has not significantly improved overall research productivity in art. The integration of AI did not result in substantial gains in productivity, likely due to the additional learning and adaptation required for implementing AI technologies. However, AI has significantly enhanced the impact of publications, leading to higher citation counts and greater academic attention. This impact is indicative of AI's role in driving research innovation and insights, although it has not translated into increased publication in high-prestige journals. The cautious approach of such journals towards multidisciplinary research may limit the dissemination of AI-enhanced art research.

The integration of AI into art research, as explored in this study, offers significant insights into the broader implications of AI's intersection with the humanities and social sciences. Art, a field deeply rooted in human creativity and cultural expression, represents a vital area of the humanities. The transformative effects observed in art research—such as enhanced interdisciplinarity, increased academic impact, and the broadening of research scope—underscore the potential for AI to similarly reshape other domains within the social sciences. However, this integration also raises critical considerations. While AI can facilitate novel methodologies and cross-disciplinary collaborations, it may also introduce challenges, such as the potential for algorithmic biases and the devaluation of traditional, qualitative approaches that are crucial for capturing the complexities of human experience. The findings of this study suggest that as AI continues to permeate the social sciences, there is a need for a careful balance between embracing technological advancements and preserving the foundational elements of humanistic inquiry. This balance is essential to ensure that AI enhances, rather than undermines, the richness and diversity of research within the humanities and social sciences.

This study has several limitations. Our research is focused on art elements within cutting-edge academic projects funded by the U.S. federal government, rather than encompassing the full scope of art research across the entire academic landscape in the U.S. This focus may not fully capture the broad impact of AI across all domains of art research. Further research is needed to explore AI's effect across a wider range of art research contexts and data sources to provide a more comprehensive understanding of its role in the field.

## SUPPLEMENTARY INFORMATION

Supplementary Notes 1, 2, and 3 can be found in a separate file within our submission. For more results and data, please refer to the Open Science Framework (OSF) link: https://osf.io/j3ysa/?view_only=9a283754d5f042a385ae88247bdfab3c.

**Supplementary Information for**

The Impact of Artificial Intelligence on Art Research: An Analysis of Academic Productivity and Multidisciplinary Integration

Yang Ding [1,2,3*]

[1] Business School, University of Edinburgh, Edinburgh EH8 9JS, United Kingdom

[2] Edinburgh Futures Institute, University of Edinburgh, Edinburgh EH3 9EF, United Kingdom

[3] Lab for Interdisciplinary Spatial Analysis, University of Cambridge, Cambridge CB3 9EP, United Kingdom

* Correspondence concerning this article should be addressed to Yang Ding (Lab for ISA, 19 Silver Street, University of Cambridge, Cambridge CB3 9EP, email: yang.ding@ed.ac.uk).

# Supplementary Note 1: Distribution of Research Projects Across NSF Directorates and Disciplines

Tables S1 and S2 show that the application of art in biology and computer science is relatively extensive, whereas its application in engineering, geosciences, and fundamental sciences is limited. The education field places a high emphasis on the application of art and explores the potential of integrating art with AI. The social sciences also show significant interest in applying art, particularly in explaining and understanding human behavior and socio-economic phenomena.

**Table S1. Distribution of Distribution of arts projects in the Biology, Computer and Information Sciences, Education and Engineering directorates.**

| Discipline | Directorate | Art | AI + Art | Non-art |
|---|---|---|---|---|
| DBI (Division of Biological Infrastructure) | BIO | 38 | 0 | 11341 |
| DEB (Division of Environmental Biology) | BIO | 25 | 0 | 26407 |
| MCB (Division of Molecular and Cellular Biosciences) | BIO | 23 | 0 | 19067 |
| IOS (Division of Integrative Organismal Systems) | BIO | 22 | 0 | 38659 |
| CCF (Division of Computing and Communication Foundations) | CISE | 26 | 1 | 19290 |
| OAC (Office of Advanced Cyberinfrastructure) | CISE | 23 | 1 | 5028 |
| CNS (Division of Computer and Network Systems) | CISE | 30 | 3 | 21070 |
| IIS (Division of Information and Intelligent Systems) | CISE | 28 | 11 | 14194 |
| EES (Directorate for Environmental Engineering and Science Excellence and Equity) | EES | 25 | 1 | 4993 |
| DGE (Division of Graduate Education) | EHR | 18 | 0 | 32963 |
| HRD (Division of Human Resource Development) | EHR | 4 | 0 | 5109 |
| DUE (Division of Undergraduate Education) | EHR | 38 | 1 | 17708 |
| DRL (Division of Research on Learning in Formal and | EHR | 38 | 8 | 10407 |

| Discipline | Directorate | Art | AI + Art | Non-art |
|---|---|---|---|---|
| Informal Settings) | | | | |
| EEC (Division of Engineering Education and Centers) | ENG | 27 | 0 | 16790 |
| CMMI (Division of Civil, Mechanical, and Manufacturing Innovation) | ENG | 24 | 0 | 30806 |
| CBET (Division of Chemical, Bioengineering, Environmental, and Transport Systems) | ENG | 21 | 0 | 30114 |
| ECCS (Division of Electrical, Communications, and Cyber Systems) | ENG | 15 | 0 | 13399 |
| EFMA (Emerging Frontiers and Multidisciplinary Activities) | ENG | 6 | 0 | 864 |
| EF (Division of Emerging Frontiers) | ENG | 4 | 0 | 987 |

**Table S2. Distribution of art projects in Geosciences, Mathematics and Physics, Social Sciences and Frontier Studies.**

| Discipline | Directorate | Art | AI + Art | Non-art |
|---|---|---|---|---|
| EAR (Division of Earth Sciences) | GEO | 24 | 0 | 31928 |
| OCE (Division of Ocean Sciences) | GEO | 21 | 0 | 23109 |
| AGS (Division of Atmospheric and Geospace Sciences) | GEO | 16 | 0 | 15041 |
| PLR (Division of Polar Programs) | GEO | 5 | 0 | 7650 |
| CHE (Division of Chemistry) | MPS | 22 | 0 | 67091 |
| DMS (Division of Mathematical Sciences) | MPS | 29 | 0 | 27150 |
| DMR (Division of Materials Research) | MPS | 27 | 0 | 25709 |
| AST (Division of Astronomical Sciences) | MPS | 19 | 0 | 8251 |
| PHY (Division of Physics) | MPS | 9 | 0 | 13957 |
| OSI (Office of Special Initiatives) | MPS | 2 | 0 | 142 |
| OIA (Office of Integrative Activities) | OIA | 30 | 1 | 1475 |
| OISE (Office of International Science and Engineering) | OISE | 24 | 1 | 28520 |
| RISE (Directorate for Research, Innovation, Synergy, and Education) | RISE | 9 | 0 | 1278 |
| SMA (Social Science Multidisciplinary Activities) | SBE | 12 | 0 | 1391 |
| NCSE (National Center for Science and Engineering Statistics) | SBE | 1 | 0 | 718 |
| BCS (Division of Behavioral and Cognitive Sciences) | SBE | 33 | 2 | 19482 |
| SES (Division of Social and Economic Sciences) | SBE | 31 | 2 | 17734 |

Frontier projects combining AI and art are primarily concentrated in the fields of computer science and education, highlighting the importance and foresight of these fields in multidisciplinary innovation research. The CISE directorate is a major force driving research that combines art and AI. The Division of Information and Intelligent Systems (IIS) stands out with 28 art-as-core projects, along with 11 frontier projects combining AI and art. This indicates a high level of interest and potential application in the field of computer science, particularly in intelligent systems and network infrastructure. In contrast, the EHR directorate emphasizes the application of art in education, particularly in the Division of Undergraduate Education (DUE) and the Division of Research on Learning in Formal and Informal Settings (DRL), each with 38 art-as-core projects. Furthermore, the DRL also has 8 frontier projects combining AI and art, reflecting the federal government's active exploration of the potential for integrating art with AI in the education field.

The distribution of art-related projects across various scientific disciplines, as detailed in Table S2, highlights the uneven integration of art within Geosciences, Mathematics and Physical Sciences, Social Sciences, and Frontier Studies. In the Geosciences (GEO) directorate, divisions such as Earth Sciences (EAR), Ocean Sciences (OCE), Atmospheric and Geospace Sciences (AGS), and Polar Programs (PLR) exhibit very limited engagement with art-related initiatives, with no projects combining AI and art. A similar pattern is observed within the Mathematics and Physical Sciences (MPS) directorate, where divisions like Chemistry (CHE), Mathematical Sciences (DMS), and Physics (PHY) have minimal involvement with art, and no AI-art projects are documented.

However, the Office of Integrative Activities (OIA) and the Office of International Science and Engineering (OISE) demonstrate a higher level of multidisciplinary research by incorporating AI-art projects, each contributing one such initiative. The Social, Behavioral, and Economic Sciences (SBE) directorate, particularly the Division of Behavioral and Cognitive Sciences (BCS) and the Division of Social and Economic Sciences (SES), shows a more significant integration of art, including several AI-art projects. This suggests a growing recognition of the potential of art in enhancing research in these fields, particularly in exploring social and economic phenomena. Overall, the data underscores the varied levels of engagement with art across disciplines, with most scientific fields displaying limited integration, while specialized offices and social sciences are more actively incorporating art, particularly in combination with AI.

## Supplementary Note 2: Clustered Keywords and TF-IDF Values from NSF-Funded Research Projects

This appendix presents a summary of the keywords and their corresponding TF-IDF values extracted from various NSF-funded research projects. The data were obtained through a text clustering analysis of project abstracts across multiple disciplines. For each cluster, we have showcased the top five keywords along with their TF-IDF values, which represent the core themes and focus areas within each cluster.

The dataset includes the following columns: the discipline name (Subject), cluster number (Cluster), the top five keywords from each cluster (Keyword 1 to Keyword 5), and the corresponding TF-IDF values (TF-IDF 1 to TF-IDF 5). These keywords and their associated weights offer valuable insights into the thematic structure and research focus within each discipline.

In total, the abstracts were clustered into 263 distinct groups across various disciplines. Given the extensive nature of this dataset, only a subset of the data is displayed in this appendix to illustrate the structure and content. The full dataset, which includes all keywords and TF-IDF values for each cluster, has been uploaded to the Open Science Framework (OSF) for further reference and analysis. To access the complete results, please visit the following link: https://osf.io/j3ysa/?view_only=9a283754d5f042a385ae88247bdfab3c. The clustering results for NSF-funded research projects combining AI and ART, as well as those focusing on traditional ART, are also available at this OSF link.

## Supplementary Note 3: Results of the impact of the integration of artificial intelligence and art on the multidisciplinary of knowledge

**Table S3. TWFE regression estimates of the impact of the integration of artificial intelligence and art on the multidisciplinary of knowledge.**

| Variables | Weighted similarity | Max similarity |
|---|---|---|
| AI | 0.507*** | 0.259*** |
|  | (0.043) | (0.036) |
| Treatment | 0.092*** | 0.043** |
|  | (0.015) | (0.013) |
| Gender dummy | 0.039*** | 0.023* |
|  | (0.011) | (0.009) |
| Training | 0.011 | 0.007 |
|  | (0.012) | (0.010) |
| Imitation isomorphism | 0.000 | 0.000 |
|  | (0.000) | (0.000) |
| Academic year | 0.016*** | 0.008*** |
|  | (0.000) | (0.000) |
| Field paper count | 0.030*** | 0.017** |
|  | (0.007) | (0.006) |
| Field citation count | -0.018** | -0.010* |
|  | (0.006) | (0.005) |
| Affiliation paper count | -0.045*** | -0.015 |
|  | (0.013) | (0.011) |
| Affiliation citation count | 0.039*** | 0.014 |
|  | (0.011) | (0.009) |
| Awarded amount to date | -0.072*** | -0.036*** |
|  | (0.003) | (0.003) |
| Constant | 0.703*** | 0.839*** |
|  | (0.043) | (0.036) |

**Note:** *** p<0.001, ** p<0.01, * p<0.05. For more detailed regression results, please refer to File Regression_results_with_ci.csv in this OSF project.